\documentclass[12pt]{article}  
\usepackage{amsmath}

\usepackage{textcomp}
\usepackage{gensymb}
\usepackage{graphicx}
\usepackage{titlesec}
\usepackage[title]{appendix}
\usepackage{authblk}

\author[1]{Hemanth Pruthvi}
\author[1]{Nagaraju K.}
\author[1]{Ravindra B.}
\author[1]{K. B. Ramesh}
\affil[1]{Indian Institute of Astrophysics, Bengaluru, India, 560034}
\providecommand{\keywords}[1]{\textbf{Keywords:} #1}

\usepackage[noabbrev,capitalize]{cleveref}
\crefname{figure}{Fig.}{Figs.}
\crefname{table}{Table}{Tabs.}
\crefname{section}{Sec.}{Sections}
\crefname{equation}{Eq.}{Eqs.}
\crefname{appendix}{Appendix}{}

\date{}
\begin{document}	
\title{Solar Spectropolarimetry of Ca \textsc{ii} 8542 \AA\ Line: Polarimeter Development, Calibration and Preliminary Observations}
\maketitle
\begin{abstract}
Chromospheric magnetic fields are of paramount importance in understanding the dynamics of energetic events in the solar atmosphere. At the Kodaikanal Solar Observatory, several polarimeters were developed in the past to study the active region magnetic fields. A new polarimeter has been developed and installed at Kodaikanal Tower-tunnel Telescope to study the active regions at Chromospheric level, in Ca~\textsc{ii} 8542~\AA\ spectral line. Design aspects of the instrument and polarimetry strategy are discussed. Telescope instrumental polarization has been revisited and possible ways to reduce it have been proposed. Telescope polarization model developed in \textsc{Zemax} to examine the analytical instrumental polarization model is discussed. The polarimeter control unit, and the software developed to operate the polarimeter are briefly described. Polarimetric calibration of the instrument, observations, corrections for instrumental polarization and the sample Stokes profiles are presented. Polarimetric accuracy and sensitivity are estimated to be better than $3\times10^{-2}$ and $3\times 10^{-3}$ respectively.

\end{abstract}

\section{Introduction} \label{sec:intro}

Magnetic field acts as a coupling between different layers and plays a central role in producing energetic events in the solar atmosphere. In the past decades, observations of the magnetic fields have become an integral part of modelling and making predictions of energetic events. Today such observations are being carried out in long term and short term by space based as well as ground based observing facilities \cite{Lagg_2015_MeasPhotChrom}. Though the photospheric magnetic field measurements are being carried out on routine basis, observations of the chromospheric magnetic fields are relatively sparse (e.g., Vector SpectroMagnetoghraph (VSM) at Synoptic Optical Long-term Investigations of the Sun (SOLIS) \cite{Keller_1998_SOLISInstru} observing in Ca II 8542 \AA\ , and Infrared Stokes Polarimeter \cite{Hanaoka_2013_NAOJ} at National Astronomical Observatory of Japan, Mitaka, observing in He I 10830 \AA\ ). Chromosphere is an important interface layer between the visible photosphere and hot corona of the solar atmosphere. Many of the energetic events occur in the solar corona. To understand the physics behind these events and possibly predict them, information on coronal magnetic field  is needed. However, because of very weak magnetic field strength at the coronal heights it is difficult to measure them. A long integration of the polarization signal is required to get any meaningful observations \cite{Lin_2004_CorMagFielMeas}. Linear and non-linear force-free magnetic field extrapolation techniques have been used to compute the coronal magnetic field from photospheric vector magnetograms \cite{Tadesse_2013_NLFFEx} but, the photospheric magnetic field is not force-free \cite{Metcalf_1995_ChromForFree}. However, chromospheric magnetograms could be more suitable to extrapolate the magnetic field to coronal heights as the field at chromosphere is close to force-free state (Ref.~\cite{Metcalf_1995_ChromForFree} and references therein).

Zeeman effect is one of the most widely used diagnostic tools to infer magnetic field in the solar atmosphere \cite{Sanchez_1992_SolObsTechInt}. Apart from splitting a spectral line, with non-zero Land\'e-g factor, into its magnetic sub-levels, Zeeman effect produces characteristic polarization across the spectral line which depends on the direction of the magnetic field with respect to the observer's line-of-sight. Measurements of spectrally resolved polarization provide information on the vector magnetic field. Two techniques that are widely used to perform such observations are 1) Spectrograph based, and 2) Tunable filter based. In the former method, polarized spectra are obtained using a combination of polarimeter and spectrograph at a given spatial position, and scanning in the spatial direction over time to construct a 2D spatial information. It may involve single or multiple slits (e.g., VSM at SOLIS \cite{Keller_1998_SOLISInstru}), or optical fibers (e.g. proposed Diffraction-Limited Near InfraRed SpectroPolarimeter at Daniel K. Inouye Solar Telescope \cite{Tritschler_2015_DLNIRSP}). 
In the latter method, tunable filters are used to sample the spectral line while obtaining the polarized images of the region of interest for that particular line position. The filters may be Fabry-Perot etalon (e.g., Imaging Spectropolarimeter at Multi Application Solar Telescope \cite{Tiwary_2017_MASTImSpecPol}), birefringent crystal or Michelson interferometer based ones, or a combination of these (e.g., Helioseismic and Magnetic Imager aboard Solar Dynamics Observatory \cite{Schou_2012_HMI}). It should be noted that both the methods incorporate a polarimeter unit to obtain the polarization information and have tradeoffs. 

In the present case, spectrograph based technique is used at one of the existing solar observational facilities: Kodaikanal 
Tower-tunnel Telescope (KTT)  hosted at Kodaikanal Solar Observatory (KSO), as it already has a High Resolution Spectrograph (HRS) as back-end. A polarimeter has been developed and installed at KTT. It is aimed towards probing the chromospheric magnetic field by carrying out spectro-polarimetric observations of Ca~\textsc{ii} 8542~\AA\ spectral line. Some of the reasons for choosing this line for diagnosing the chromospheric magnetic field are \cite{Uitenbroek_2010_ForModelChrom} :

\begin{itemize}
	\item The line has continuous sensitivity from photosphere to middle chromosphere: line wings form in the photosphere and core at an height of $\approx$1500 km \cite{Lagg_2015_MeasPhotChrom}. 
    \item It has good sensitivity to magnetic fields with Land\'e \textit{g}-factor of 1.10.
    \item Inversion techniques are well developed and tested.
    \item Existing Silicon based detectors can be used without much demand.
\end{itemize}

In the next section, description of the existing system is given along with the design and development aspects of the polarimeter that is tailored to it. Instrumental polarization of the telescope and possible solutions to mitigate it are also discussed. On-site calibration of the instrument, preliminary observation and data reduction are discussed in \cref{sec:calobs}. Instrument application and future work are summarized in \cref{sec:summary}.

\section{Design aspects}

\begin{figure}
\centering
\includegraphics[width=150mm]{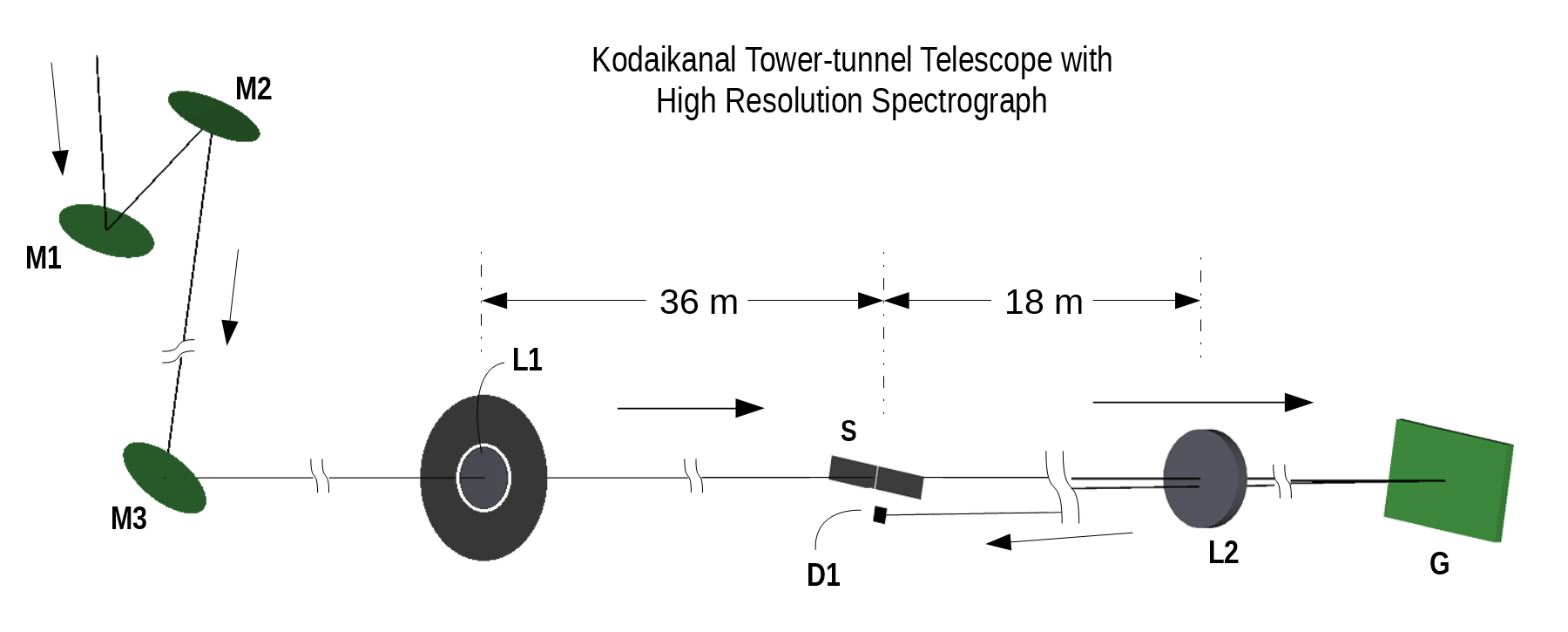}
\caption{A schematic of the Kodaikanal  tower tunnel telescope and spectrograph unit. M1, M2 and M3 are plane mirrors of the Coelostat system. L1 is the objective lens that forms the image on slit denoted by S. L2 is collimating and imaging lens of the spectrograph, and G is the reflecting grating. D1 is the science detector which is placed just below the slit.}
\label{fig:ktthrs}
\end{figure}

At KTT, a 2-mirror Coelostat placed in the tower is used as a light feeding system. Primary mirror (M1) of the Coelostat tracks the Sun and the secondary mirror (M2) guides the light vertically down. This beam is fed into the horizontal tunnel by a fold mirror (M3) with $45\degree$ orientation. M3 directs the beam on to an air-spaced doublet lens (L1) which is the objective. Slit based HRS is located at the focal plane of L1. The spectrograph has Littrow configuration, with another air-spaced doublet (L2) acting as a collimator as well as an imager. Blazed reflection grating (G) is the dispersing element in the HRS. Spectrum is focused just below the slit where the science detector (D1) is placed. The spectrograph is a fixed setup \cite{Bappu_1967_SolPhysAtKKL}. A schematic of KTT along with the spectrograph setup is shown \cref{fig:ktthrs}.

Previously, multiple polarimeters were developed aiming to measure the vector magnetic fields at photospheric \cite{Sankar_2002_MeasSolVecMagFieKTT} as well as at chromospheric heights \cite{Nagaraju_2008_Halpha}. First with the single beam \cite{Bala_1985_TheoInstruPolKTT, Sankar_2002_MeasSolVecMagFieKTT} and later with dual beam \cite{Nagaraju_2007_EffModSchemeDBPol, Nagaraju_2008_PerfDBPolKTT} setup. Some of them suffered from spurious seeing induced cross-talks while others were too slow to measure the Stokes parameters for the whole of a sunspot region. Considering these drawbacks, a new polarimeter is designed with following goals:
\begin{itemize}
	\item Improved cadence,
	\item Spatial scanning mechanism which is an integral part of the polarimeter,
    \item Reduced instrumental polarization,
    \item Provision for context imaging and auto-guiding.
    
\end{itemize}
Important specifications of the KTT system are listed in \cref{tab:telespecs}.
\begin{table}
\centering
\caption{Specifications of the KTT, spectrograph and detector.}
\begin{tabular}{ll}
	\hline\noalign{\smallskip}
	Front-end specifications & \\
	\noalign{\smallskip}\hline\noalign{\smallskip}
	M1, M2, M3 aperture & 60 cm \\
	L1 aperture & 38 cm \\
	F-ratio & 96 \\
	Image plate scale & 5.5 arcsec/mm \\
	\noalign{\smallskip}\hline\noalign{\smallskip}
	Spectrograph specifications & \\
	\noalign{\smallskip}\hline\noalign{\smallskip}
	Spectrograph configuration & Littrow \\
	L2 aperture & 20 cm \\
	L2 focal length & 183 cm \\
	Grating groove density & 600 lines/mm \\
	Blaze angle & 55$\degree$ \\
	\noalign{\smallskip}\hline\noalign{\smallskip}
	Detector specifications & \\
	\noalign{\smallskip}\hline\noalign{\smallskip}
	Type & CCD \\
	Format & 2048 $\times$ 2048 \\
	Pixel size & 13.5 $\mu$m \\
	Sensor area & 27.6 mm $\times$ 27.6 mm \\
	Quantum Efficiency & 50\% @ 8542 \AA \\
	Readout clock & 3 MHz \\
	\noalign{\smallskip}\hline
\end{tabular}
\label{tab:telespecs}
\end{table}


\subsection{Polarimetry} \label{ssc:theory}

The Stokes parameters are the full set of measurable quantities which completely describe a general state of polarization of fully or partially polarized light. They are mathematically expressed as column vector $\Vec{S} = [I, Q, U, V]^T$, with $^T$ denoting transpose operation. The detectors that are being used for polarimetry in visible and infrared range are only sensitive to the total intensity:$I$. Hence, the polarization has to be modulated (encoded) into intensity variation in a known way such that the state of polarization can be derived using these intensity measurements. The change in state of polarization due to a physical element is expressed using a $4 \times 4$ matrix called Mueller matrix. As only the intensity can be directly measured, only first row of the Mueller matrix is relevant for modulation. To obtain all the four Stokes parameters, at least four intensities and corresponding Mueller matrices' first rows should be known. All such intensities can be expressed in matrix form as shown in \cref{eq:modmat}.

\begin{equation}
\Vec{I_{out}} = 
\begin{bmatrix}
I^1_{out} \\ I^2_{out} \\ \vdots \\ I^n_{out}
\end{bmatrix} = 
\begin{bmatrix}
1 & m^1_{01} & m^1_{02} & m^1_{03} \\
1 & m^2_{01} & m^2_{02} & m^2_{03} \\
\vdots & \vdots & \vdots & \vdots \\
1 & m^n_{01} & m^n_{02} & m^n_{03} \\
\end{bmatrix}
\begin{bmatrix}
I_{in} \\ Q_{in} \\ U_{in} \\ V_{in}
\end{bmatrix} = 
\Vec{O}.\Vec{S_{in}}.
\label{eq:modmat}
\end{equation}

$\Vec{O}$ is called modulation matrix and its rank should be minimum 4 to measure four Stokes parameters \cite{DelToro_2000_OptModDemodSolPol}. It is shown in a generalized form for $n$ number of modulations in \cref{eq:modmat}. Each row of $\Vec{O}$ is normalized to its first element. The modulation matrix is determined from the calibration data. Light with known state of polarization ($\Vec{S_{calin}}$) is sent through the polarimeter and intensity outputs are measured for all modulations. This is repeated for at least four different known states of polarization and all the intensities are recorded ($\Vec{I_{calout}}$). From these intensity measurements, modulation matrix can be calculated as

\begin{equation}
\Vec{O} = 
\Vec{I_{calout}}.\Vec{S_{calin}^T}.\Vec{(S_{calin}.S_{calin}^T)^{-1}}.
\label{eq:modmatfromcal}
\end{equation} 

Once the modulation matrix is known, its corresponding efficient demodulation matrix is calculated \cite{DelToro_2000_OptModDemodSolPol} as

\begin{equation}
\Vec{D} = 
\Vec{(O^T.O)^{-1}.O^T}.
\label{eq:effdemod}
\end{equation}

This is applied to the modulated intensity measurements to obtain the actual input Stokes parameters.

\begin{equation}
\Vec{S_{in}} = 
\Vec{D.I_{out}}.
\label{eq:demod}
\end{equation}

A balanced scheme comprising of four stepped modulations is used for this polarimeter. It is chosen with the objective of keeping observing time to a minimum. Simplest four step modulation scheme with equal weights to Stokes Q, U, and V is taken to be modulation matrix, resulting in having equal $efficiency$ for measuring $Q$, $U$ and $V$. Modulation matrix corresponding to this scheme is given in \cref{eq:thmodmatBal}.

\begin{equation}
\Vec{O_{bal}} = 
\begin{bmatrix}
1 & \mp0.577 & \pm0.577 & \mp0.577 \\
1 & \pm0.577 & \mp0.577 & \mp0.577 \\
1 & \mp0.577 & \mp0.577 & \pm0.577 \\
1 & \pm0.577 & \pm0.577 & \pm0.577 \\
\end{bmatrix}.
\label{eq:thmodmatBal}
\end{equation}

\subsection{Polarimeter setup} \label{ssc:optdes}

A combination of Quarter-Wave Plate (QWP) and Half-Wave Plate (HWP) is used for the modulation. QWP and HWP position angles to implement the balanced modulation scheme are listed in \cref{tab:modangles}. Polarizing Beam Displacer (PBD) is used as analyzer to separate orthogonal linear states of polarized light as this gives significant advantage in terms of reduced seeing induced cross-talk over single beam polarizers \cite{Lites_1987_SeeIndCrossTalk}. 

\begin{figure}
\centering
\includegraphics[width=150mm]{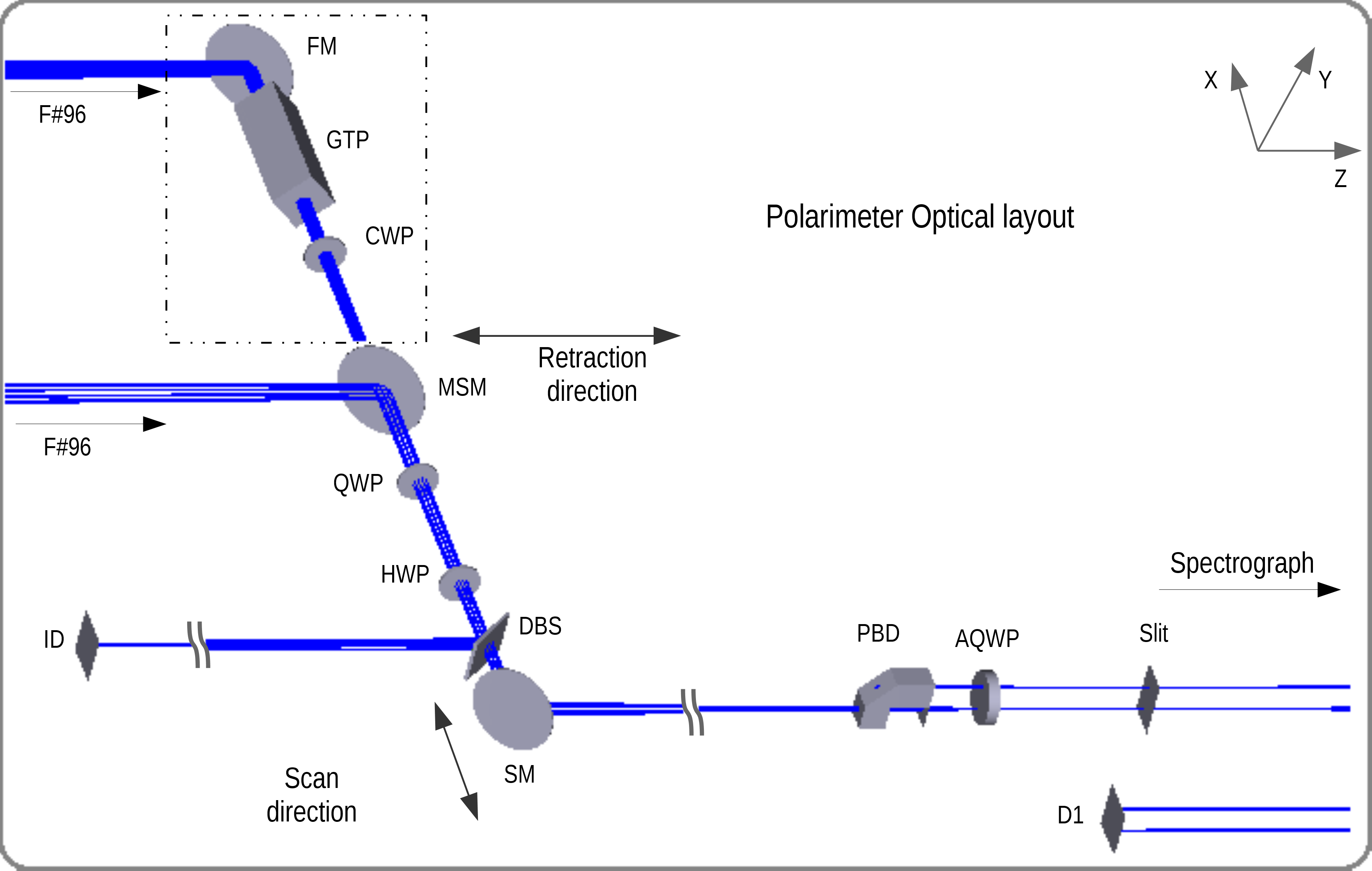}
\caption{Optical layout of the polarimeter setup at KTT. In observation mode, light from the telescope is fed to the modulation unit by a folding mirror (MSM). It passes through quarter wave plate (QWP), half wave plate (HWP), and dichroic beam splitter (DBS) that transmits light of wavelength above 600 nm. It is then reflected by another fold mirror (SM), and passes through polarizing beam displacer (PBD) that splits the beam in to two with orthogonal states of polarization, followed by achromatic quarter wave plate (AQWP). In calibration mode, light from the telescope is fed to the calibration unit by yet another folding mirror (FM). It then passes through Glam-Thompson polarizer (GTP) and quarter wave plate (CWP) producing light with known state of polarization. In this mode MSM is retracted and the light is allowed to passes through modulation unit and so on.}
\label{fig:optlay}
\end{figure}

Unlike in the previous polarimeter setup \cite{Nagaraju_2007_EffModSchemeDBPol}, this polarimeter is placed perpendicular to the optical path from objective to slit. This configuration has been worked out keeping in view of two requirements. One is to have an integrated scanning system and the other is to have an auto-guiding system. The optical layout of the polarimeter setup at KTT is shown in \cref{fig:optlay}. The incoming beam is directed through the polarimeter using a plane mirror placed inclined at $\approx45\degree$. A natural consequence of this mirror is to nullify the instrumental polarization introduced by M3, which is discussed in detail in \cref{sec:instrupol}. A Dichroic Beam Splitter (DBS), Scanning Mirror (SM) are placed between the modulation unit and analyzer. The DBS is a long pass filter with cut-off at 600 nm. The reflected beam is captured by a second detector (ID) for the purposes of context imaging and auto-guiding. The transmitted beam is reflected by SM on to the slit. SM moves along the direction of the incident beam and consequently image is moved across the slit. Although this does cause the image focal plane to shift, its effect is negligible considering the depth of focus provided by $f/96$ beam. 

\begin{table}
\centering
\caption{ List of QWP and HWP position angles to implement the selected balanced modulation scheme.}
\begin{tabular}{lll}
	\hline\noalign{\smallskip}
	No. & QWP position & HWP position \\
	\noalign{\smallskip}\hline\noalign{\smallskip}
	1 & -22.5$\degree$ & 42.6$\degree$ \\
	2 & -22.5$\degree$ & 69.9$\degree$ \\
	3 & 22.5$\degree$ & 47.4$\degree$ \\
	4 & 22.5$\degree$ & 20.1$\degree$ \\
	\noalign{\smallskip}\hline\noalign{\smallskip}
\end{tabular}
\label{tab:modangles}
\end{table}

A calibration unit consisting of a Glan-Thompson Polarizer (GTP) and a Calibration Wave Plate (CWP) is used to generate the known state of polarization. GTP polarizes the light to a very high degree ($\approx 10^5$) and a zero-order QWP is used as CWP.  An Achromatic Quarter-Wave Plate (AQWP) is used to convert two linearly polarized outputs into circularly polarized outputs, so that response of the grating is same for both the beams. In this mode, MSM is retracted completely so that light from the calibration unit may pass through the modulation unit.

\begin{table}
\centering
\caption{Specifications of the instrument and settings of the spectrograph.}
\begin{tabular}{ll}
	\hline\noalign{\smallskip}
	Specification & Value\\
	\noalign{\smallskip}\hline\noalign{\smallskip}
	Wavelength of interest & 8542 \AA  \\
	Order & 2 \\
	Slit width & 110 $\mu$m \\
	Detector binning & 2$\times$2 \\
	Linear dispersion & 5.3 m\AA/pixel \\
	Slit equivalent wavelength extent & 42 m\AA \\
	Spectral resolution &  64 m\AA \\
	Wavelength coverage & 10.8 \AA \\
	Field of View & 60 arcsec \\
	\noalign{\smallskip}\hline\noalign{\smallskip}
	Retardance of the calibration wave plate & 0.248$\lambda$ \\ 
	PBD entrace size & 12 mm \\
	Extent of the scan & 23 mm \\
	Wave plate rotation speed & 15 rpm \\
	Scanning speed & 1 mm/s \\
	Scan step & 110 $\mu$m \\
	\noalign{\smallskip}\hline\noalign{\smallskip}
\end{tabular}
\label{tab:desspecs}
\end{table}


\subsection{Instrumental Polarization} \label{sec:instrupol}

Instrumental polarization of the Coelostat was recognized as a significant hurdle, and was modelled and measured for over 100 years \cite{StJohn_1909_PolEffCoelMirr, Hale_1912_PolPhenCoelTele, Bumba_1962_PolSolSpecCoelMirr, Capitani_1989_PolPropZeissCoel, Demidov_1991_PolCharJenschCoel}. Earlier works to develop polarimeter for KTT also included developing a theoretical model for the instrumental polarization \cite{Bala_1985_TheoInstruPolKTT}. This model was used in subsequent polarimeters \cite{Sankar_1999_MeasInstruPolCoelKTT, Nagaraju_2010_SunSpotMag}. The new model, which essentially is an extension of the previous model, is developed aiming to find ways to reduce the instrumental polarization, and it is discussed in detail further. \textsc{Zemax} model of the Coelostat is also created and the resulting output instrumental polarization values are compared. All the mirrors have unprotected Aluminium coating, and its complex refractive index (RI), at 8542~\AA\ wavelength, is taken to be $2.16-7.18\iota$ \cite{McPeak_2015_AlRI}.

\begin{figure}
    \centering
    \includegraphics[width=150mm]{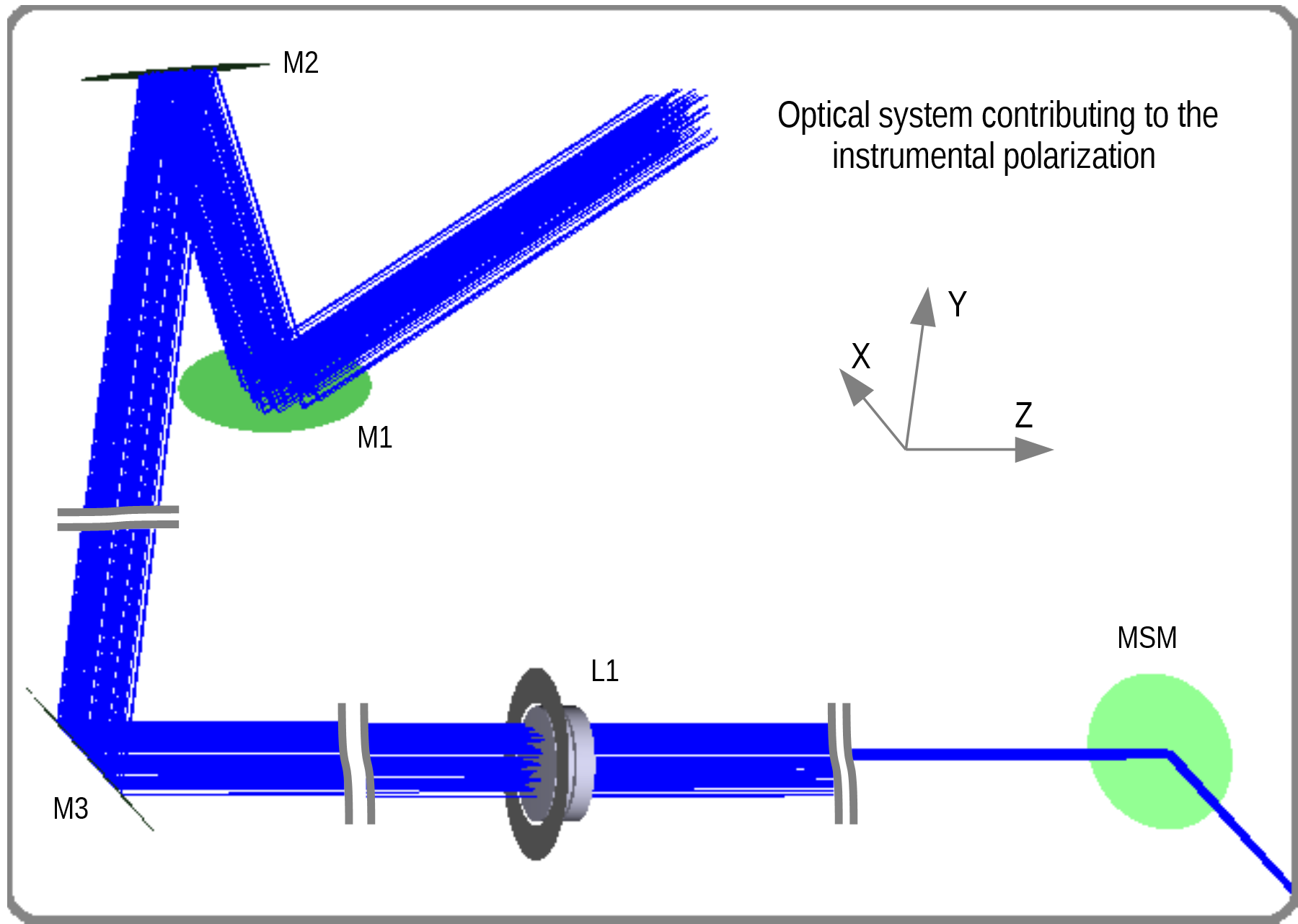}
    \caption{Optical system contributing to the instrumental polarization: M1, M2, M3, L1 and MSM.}
\label{fig:instrupolsys}
\end{figure}

In the present model, Coelostat Mueller matrix is treated to be a product of two Mueller matrices: time independent and time varying. Time independent part consists of effects of the tertiary mirror and MSM. By virtue of the design of the polarimeter, the configuration of time independent part is created in such a way that the Mueller matrix is close to identity \cite{Cox_1976_CompInstruPolIncMirr}. Results of the theoretical model and the \textsc{Zemax} model are given in \cref{eq:coeltimeind}. The model and simulation are detailed in \cref{app:instrupol}.

\begin{subequations}
\begin{equation}
\Vec{M_{CTI-Th}} = 
\begin{bmatrix}
 1.0000 &  0.0006 & -0.0007 & -0.0001 \\
 0.0005 &  0.9999 &  0.0138 &  0.0025 \\
-0.0007 & -0.0138 &  0.9999 &  0.0019 \\
-0.0001 & -0.0025 & -0.0019 &  1.0000 \\
\end{bmatrix},
\end{equation}

\begin{equation}
\Vec{M_{CTI-Si}} = 
\begin{bmatrix}
 1.0000 &  0.0016 & -0.0007 & -0.0001 \\
 0.0016 &  0.9999 &  0.0138 &  0.0025 \\
-0.0008 & -0.0137 &  0.9999 &  0.0052 \\
-0.0001 & -0.0024 & -0.0050 &  0.9999 \\
\end{bmatrix}.
\end{equation}
\label{eq:coeltimeind}
\end{subequations}

The disagreement of certain terms of the Mueller matrices can be attributed to the fact that the theoretical model only considers two mirrors in collimated beam where as the \textsc{Zemax} model is based on real configuration of the system i.e., it also accounts for the objective lens after M3 and the effects of $f/96$ beam on the MSM.

As for the time varying part, the instrumental polarization is very high in the morning and evening, and gradually decreases with decreasing magnitude of hour angle ($HA$). Mueller matrices obtained from theoretical model and simulation for $HA = -60\degree$, $DEC = 0\degree$ are given in \cref{eq:coeltimevar}, and they agree to the extent of $10^{-4}$.

\begin{subequations}
\begin{equation}
\Vec{M_{CTV-Th}} = 
\begin{bmatrix}
 1.0000 &  0.0949 & -0.0063 & -0.0018 \\
-0.0946 & -0.9852 &  0.1677 &  0.0334 \\
-0.0096 & -0.1697 & -0.9288 & -0.3156 \\
-0.0018 & -0.0220 & -0.3166 &  0.9435 \\
\end{bmatrix},
\end{equation} 
\begin{equation}
\Vec{M_{CTV-Si}} =
\begin{bmatrix}
 1.0000 &  0.0949 & -0.0063 & -0.0018 \\
-0.0946 & -0.9852 &  0.1677 &  0.0334 \\
-0.0097 & -0.1697 & -0.9288 & -0.3156 \\
-0.0018 & -0.0220 & -0.3166 &  0.9435 \\
\end{bmatrix}.
\end{equation}
\label{eq:coeltimevar}
\end{subequations}

Using this theoretical model it is calculated that if the primary mirror position is switched, from east to west ($HA<-45\degree$) or west to east ($HA>45\degree$), the instrumental polarization of the system reduces drastically (similar results were also noted earlier \cite{Bachman_1984_RedInstruPolCoelTele}). The $HA$ limitation is posed by the rigid opto-mechanical design of the KTT Coelostat. For the same $HA = -60\degree$ and $DEC = 0\degree$, Mueller matrices for the proper configuration and switched configuration are given in \cref{eq:coelcomp}.

\begin{subequations}
\begin{equation}
\Vec{M_{C4-Prop}} = 
\begin{bmatrix}
 1.0000 &  0.0944 & -0.0055 & -0.0017 \\
-0.0942 & -0.9875 &  0.1540 &  0.0314 \\
-0.0090 & -0.1562 & -0.9316 & -0.3142 \\
-0.0017 & -0.0193 & -0.3152 &  0.9441 \\
\end{bmatrix},
\end{equation} 

\begin{equation}
\Vec{M_{C4-Swit}} =
\begin{bmatrix}
 1.0000 &  0.0126 &  0.0071 & -0.0001 \\
-0.0136 & -0.9871 & -0.1593 &  0.0168 \\
-0.0050 &  0.1583 & -0.9862 & -0.0452 \\
-0.0001 &  0.0238 & -0.0420 &  0.9987 \\
\end{bmatrix}.
\end{equation}
\label{eq:coelcomp}
\end{subequations}

\begin{figure}
    \centering
	\includegraphics[width=150mm]{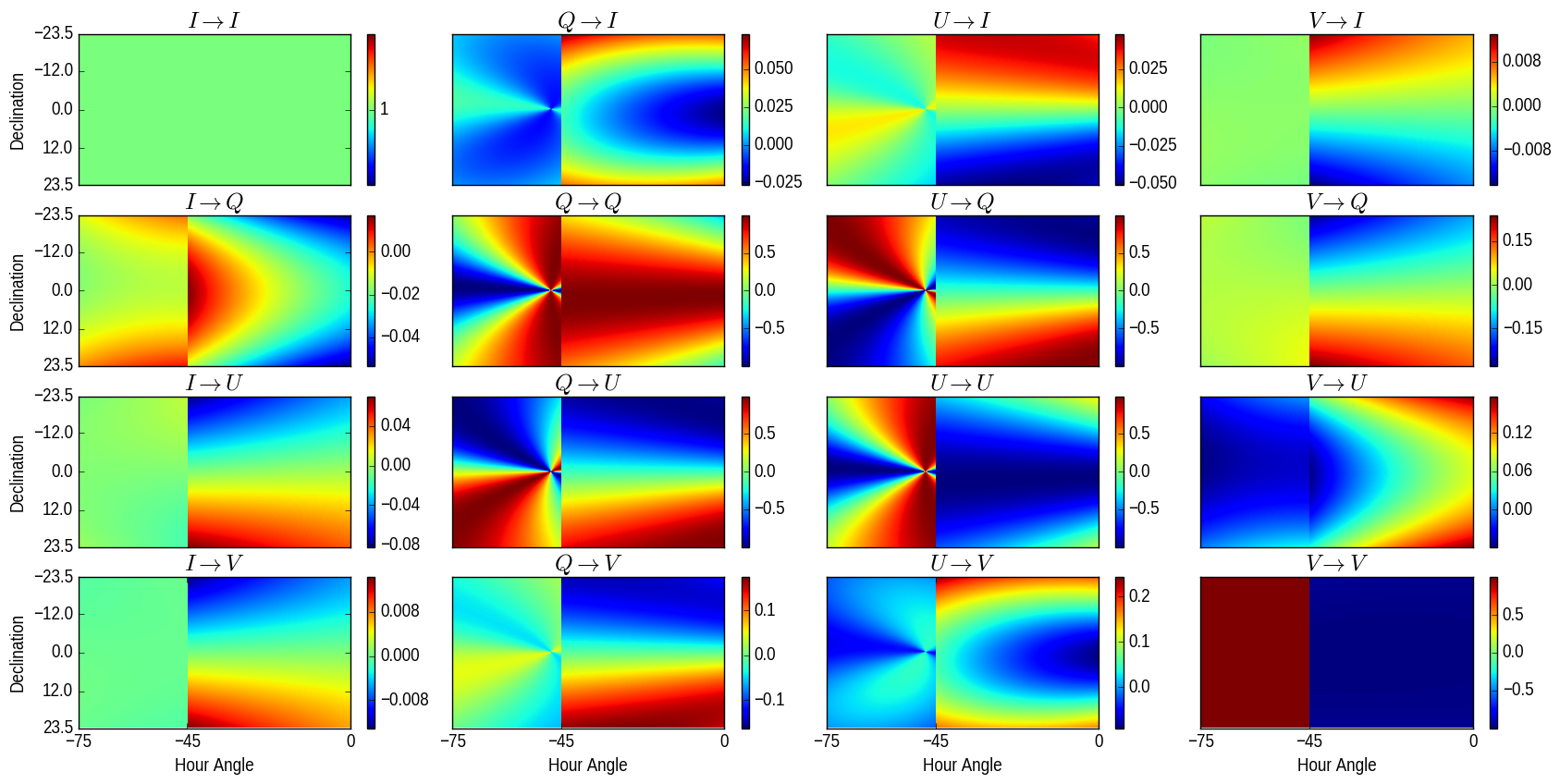}
	\caption{Mueller matrix corresponding to the system before the polarimeter.}
	\label{fig:instrupol}
\end{figure}

\begin{figure}
    \centering
	\includegraphics[width=150mm]{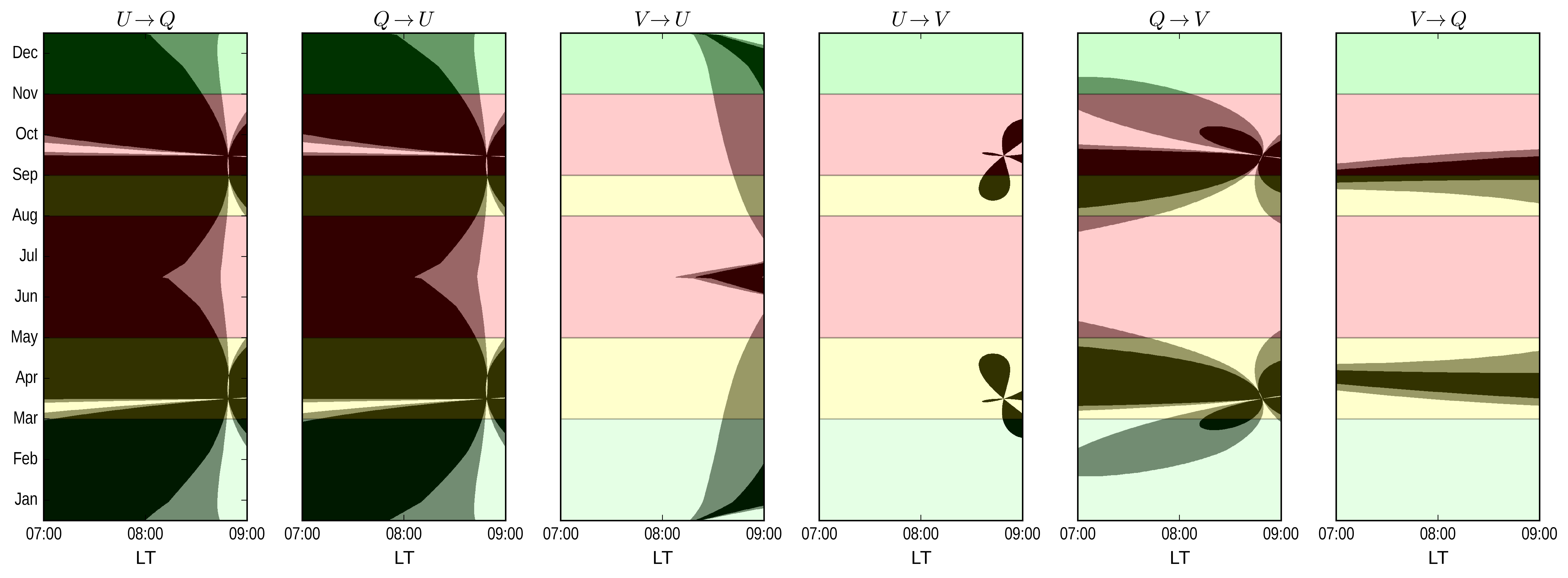}
	\caption{Illustration of effect of change in configuration and introduction of MSM on the key terms of instrumental polarization. Ratio of absolute values are rounded and plotted against local time on X-axis and month on Y-axis. Black colour indicates degradation, gray indicates improvement up to 2 times and white indicates improvement more than 2 times. Horizontal bands of colour indicate  sky conditions. Green indicates good conditions, red indicates mostly cloudy conditions and yellow indicates the rest of intermediate conditions.}
	\label{fig:instrupolimp}
\end{figure}

The final Mueller matrix of the entire system with possible improvements, as a function of $HA$ and $DEC$ is mapped in \cref{fig:instrupol}. The diagonal elements of the Mueller matrix correspond to the transmission. Non-diagonal elements of first row correspond to the cross-talks from $Q$, $U$ and $V$ to the total intensity. These values are usually very low and considering the degree of polarization ($\approx 0.2$), the cross-talk becomes negligible. Non-diagonal elements of first column correspond to the cross-talks from total intensity to $Q$, $U$ and $V$. They manifest as constant offsets in normalized Stokes $Q$, $U$ and  $V$ profiles, and to a large extent it is correctable by subtracting respective continuum polarization level from each profile. Rest of the elements correspond to cross-talks from one state of polarization to the other. \cref{fig:instrupolimp} depicts the comparison of these elements in usual conditions and after changing the configuration. Considering the Coelostat design, configuration can be changed only for the duration corresponding to $-75\degree < HA < -45\degree $. Hence, comparison for only to those results are shown. The same results can be extended for $75\degree > HA > 45\degree$. Black shade indicates degradation, gray indicates up to 2 times improvement and white indicates more than 2 times improvement. Horizontal bands of colour indicate observing conditions at KSO in an approximate manner. Green indicates good conditions (peak winter), red indicates mostly cloudy conditions (monsoons) and yellow indicates the rest of intermediate conditions.

As it is evident, cross-talks corresponding to $Q \rightarrow V$, $U \rightarrow V$, $V \rightarrow Q$ and $V \rightarrow U$ have reduced significantly. These cross-talks manifest as distortions in the Stokes $Q$, $U$ and $V$ profiles hence, any residuals after corrections can create hurdles in inverting those profiles to obtain magnetic field information. However, $Q \rightarrow U$ and $U \rightarrow Q$ cross-talks imply the rotation of coordinates and residuals manifest as error in estimation of azimuth angle of the transverse magnetic field component. Hence, it was decided to adopt the improved Coelostat configuration, and MSM was placed before the modulation unit.

In the present model, RI value was taken from existing literature. But, this is only for the purposes of comparing the model and simulation. For accurate estimation of instrumental polarization few more things must be considered. The mirrors are coated at in-house facility with bare Aluminium. After evaporation an oxide layer is formed that protects the coating to certain extent \cite{Sankar_1996_EllipsCoelBabComp}. The oxide layer is also included in the present model that is to be fitted to the observations. Profiles of five cross-talk elements of telescope Mueller matrix shall be obtained from the observations spanning few hours (\cref{ssc:obs}). RI of the coating, oxide layer thickness and folding angle of MSM are the variables that need to be estimated from the fitting. Using these values, a more accurate estimate shall be computed for the instrumental polarization.


\subsection{Controller and Software} \label{sec:soft}

\begin{figure}
    \centering
	\includegraphics[width=150mm]{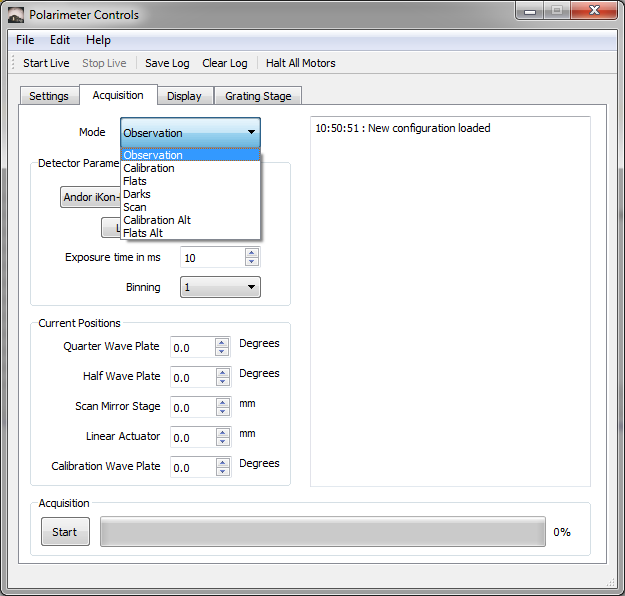}
	\caption{Graphical User Interface (GUI) developed using Python to control the whole polarimeter unit, science detector and the grating stage.}
	\label{fig:gui}
\end{figure}

The polarimeter is operated using a custom made controller, based on micro-controllers. It receives commands from the computer via a serial port and executes them. Three motorized rotation mounts for the wave plates and two motorized linear stages for the mirrors, with home positioning sensors, are controlled. Stepper motor with micro-stepping drives are used for the rotation and translation. A custom software with GUI is also developed to command the controller and spectrograph grating, and to acquire the images from the detector. The software is written in Python. Python library of Micro-Manager \cite{Micro-Manager} is used for operating the detector. User can select the mode of acquisition and other parameters for the observations, and all the actions are logged for the diagnostics purposes. \cref{fig:gui} shows a snapshot of the graphical user interface of the control software.

\section{Observations and Calibration} \label{sec:calobs}

\begin{figure}
    \centering
	\includegraphics[width=150mm]{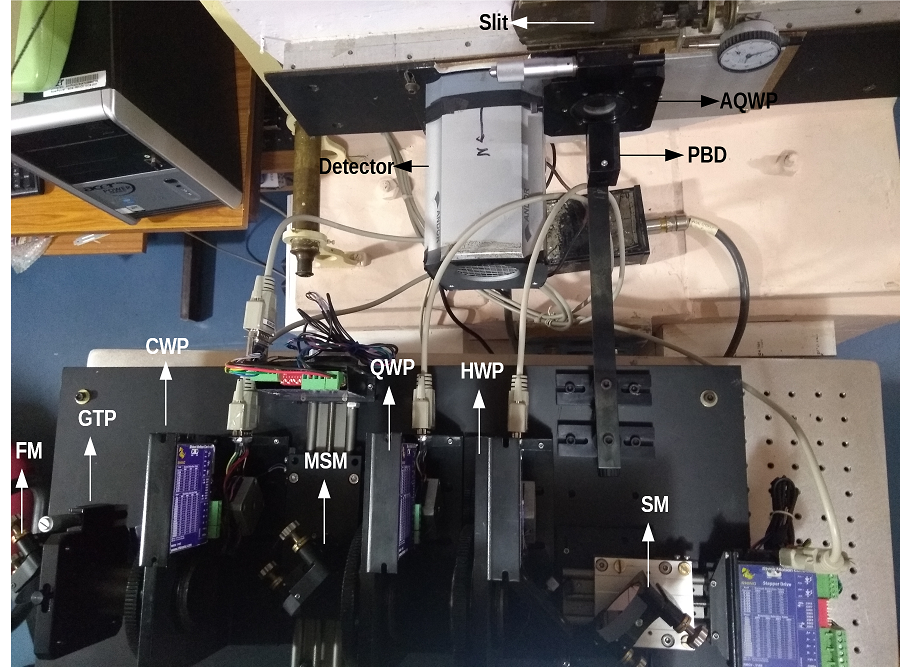}
	\caption{Instrument installed at KTT.}
	\label{fig:polinstall}
\end{figure}

\cref{fig:polinstall}  shows a top view of the polarimeter installed at the back-end of KTT. On-site calibration and preliminary observations were carried out and briefly presented here.

\subsection{Data Acquisition \& Processing} \label{ssc:dataacq}

Software is equipped to control the acquisition of four kinds of data: dark frames, flat frames, calibration frames and observation frames. Additional modes of operation are also possible but it is not relevant to current context. 
Calibration frames consist of $m\times n$ files, $m$ being the number of modulations and $n$ being the number of known input states of polarization. A minimum of four distinct known input states should be used. Here, a total of 24 known states were generated using the calibration unit, and polarization measurement was performed. As mentioned in \cref{ssc:optdes}, these states were produced by rotating the CWP in equal steps ($7.5\degree$) totaling 180 degrees. Observation frames consist of $m \times k$ where $k$ is the number of slit positions across the image.

Dark as well as flat frame acquisitions require manual intervention. Dark frames were acquired by simply closing the slit.
Flat frames were acquired by moving the disk center of the solar image randomly over the slit. Two kinds of flats were acquired: 1) main flat frames with grating position being identical to the one in observation mode, 2) fringe flat frames that were taken with near by spectral continuum being recorded on the detector. A number of python scripts have been developed to  reduce the raw data and key processes are outlined below.

 
Master dark synthesis was started by averaging all the dark frames. Note that these dark frames also account for the background. Median filter with $3\times3$ window was applied to mean dark frame to remove any ``hot/cold" pixels. They are isolated pixels with unusually high/low counts. The result was saved as master dark.  

\begin{figure}
	\centering
	\includegraphics[width=150mm]{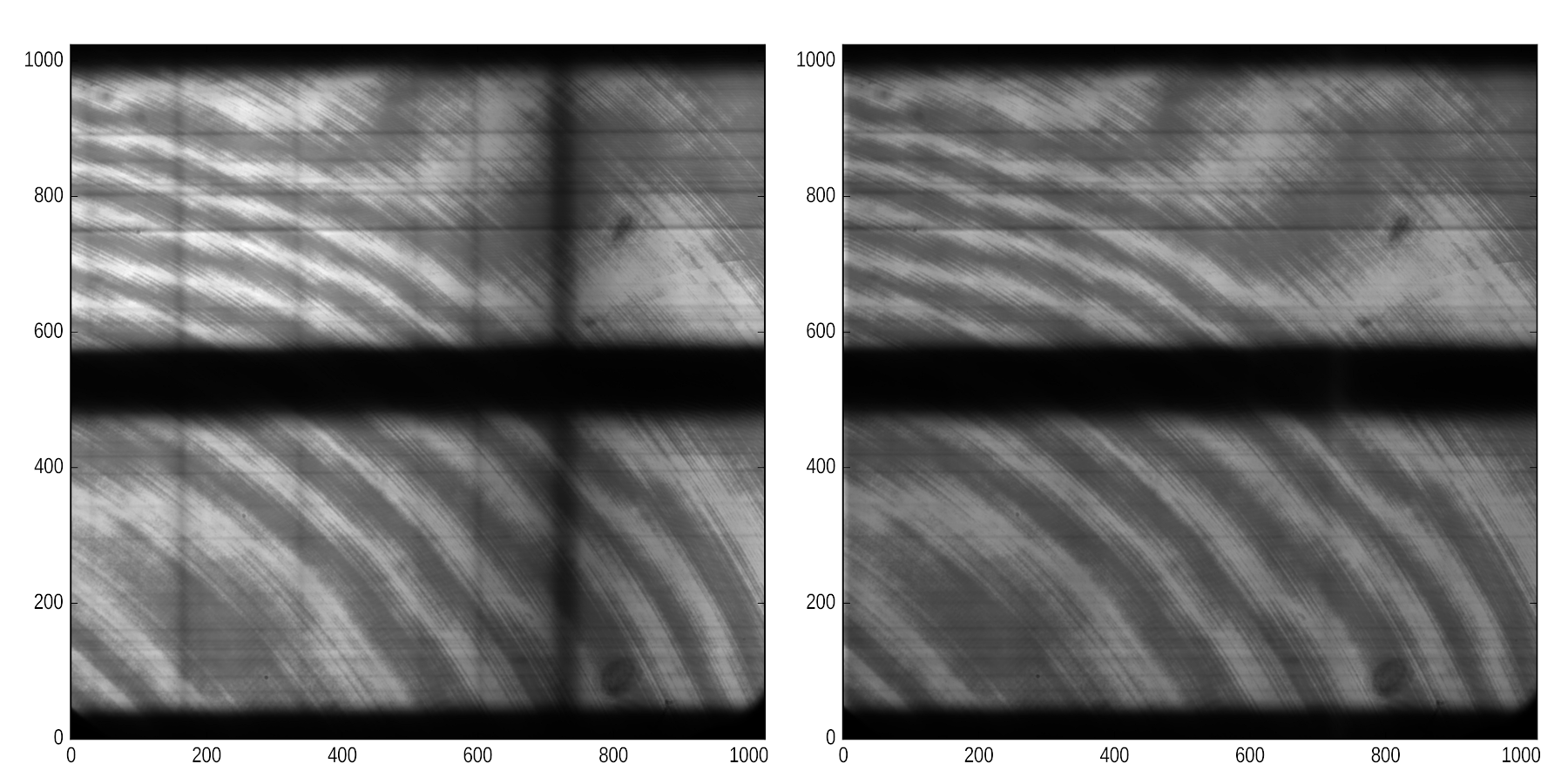}
	\caption{Left: raw flat after acquisition, right: master flat.}
	\label{fig:flats}
\end{figure}

Master flat synthesis also starts with averaging the main flat frames and subtracting master dark from it. The procedure described in Ref.~\cite{Wohl_2002_PrecRedSolSpecCCD} was followed with a few modifications that are described in \cref{app:flatmaster}. \cref{fig:flats} shows a raw flat image. The presence of high contrast fringes cannot be missed. This is due to the etaloning effect in the detector's sensor. As the sensor was optimized for blue wavelength (400 nm), a back-thinned and back-illuminated chip was used. This causes higher wavelength ($>$ 700 nm) radiation to undergo multiple internal reflections before getting absorbed. These fringes would affect the process to align the spatial and spectral axes, and median spectral line profile that is required to create the master flat. In order to overcome these setbacks, additional procedure was followed to reduce contrast of the fringes (\cref{app:flatmaster}). Row and Column shift values required to align the images, and spectral line profile from averaged disk center were also obtained in the process. They were saved as level-1 data, and to be used in further reduction.

\begin{figure}
	\centering
	\includegraphics[width=150mm]{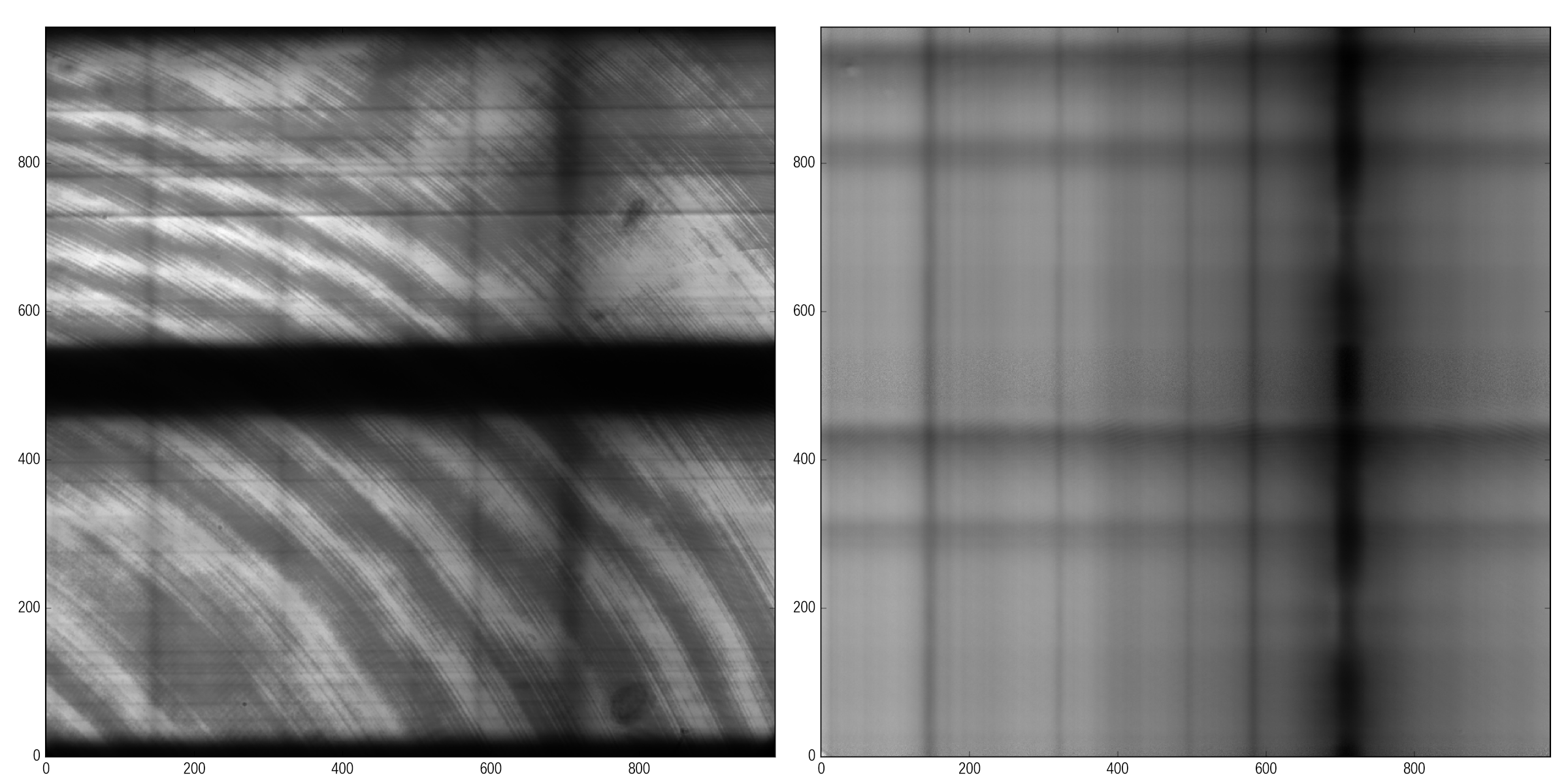}
	\caption{Left: raw data frame, right: corrected data frame.}
	\label{fig:rawcorr}
\end{figure}

Calibration and observation frames were corrected for background by subtracting the master dark. Row and column shift values were used to fine tune the orientation of spectra. It was then divided by the master flat. Frames corresponding to each modulation were stacked together and saved as ``level-1" calibration data or observation data, whichever applies. \cref{fig:rawcorr} shows an observation frame in raw state and after correction.

\subsection{Modulation matrix} \label{ssc:modmat}

\begin{figure}
    \centering
	\includegraphics[width=150mm]{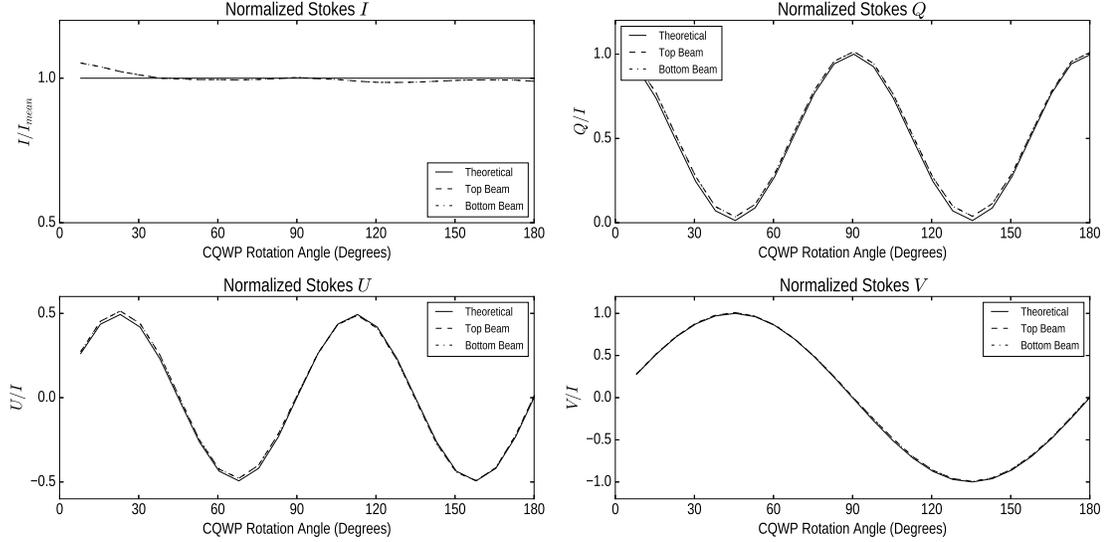}
	\caption{Demodulated PBD output for balanced modulation scheme: solid line is theoretical, dashed line is for the top beam and dash-dotted line for the bottom beam. Maximum difference between theoretical and measured curves is $\approx0.03$ (for $Q$) or less.}
	\label{fig:demodbal}
\end{figure}

Level-1 calibration data were used to calculate the modulation matrix. A Region of Interest(RoI) away from the solar spectral lines was selected in both top and bottom beams. Intensity in this RoI was integrated and modulation curves were obtained. The procedure described in \cref{ssc:theory} was followed to obtain the modulation matrix. As MSM and DBS also fall in the path of beam, measured modulation matrix would deviate from the expected one i.e., one that was calculated for the combination of QWP and HWP. Next, optimum demodulation matrix was calculated and applied to the original calibration data. By doing so, Stokes parameters of the known input states were retrieved and compared with the supposed input values. This process was iteratively performed to identify and exclude the outliers, and retrieve the position angle offset for CWP. \cref{fig:demodbal} shows the retrieved input Stokes parameters, after modulating and subsequently demodulating the input. Modulation matrices corresponding to the top and bottom beams are given by \cref{eq:modtopbot}, and demodulation matrices are given by \cref{eq:demodtopbot}.

\begin{subequations}
    \begin{equation}
        \Vec{O_{top}} = 
        \begin{bmatrix}
			1.0 &  0.560 &  0.450 &  0.589 \\         
			1.0 & -0.532 &  0.599 &  0.557 \\         
			1.0 &  0.488 &  0.604 & -0.513 \\         
			1.0 & -0.577 & -0.508 & -0.612 \\
        \end{bmatrix}
        \pm
        \begin{bmatrix}
			0 &  0.006 &  0.005 &  0.003 \\
			0 &  0.004 &  0.004 &  0.003 \\
			0 &  0.006 &  0.007 &  0.003 \\
			0 &  0.004 &  0.005 &  0.003 \\
        \end{bmatrix},
    \end{equation}
    \begin{equation}
        \Vec{O_{bot}} = 
        \begin{bmatrix}
			1.0 & -0.591 &  0.528 & -0.570 \\         
			1.0 &  0.523 & -0.537 & -0.570 \\         
			1.0 & -0.553 & -0.567 &  0.577 \\       
			1.0 &  0.508 &  0.538 &  0.571 \\  
        \end{bmatrix}
        \pm
        \begin{bmatrix}
			0 &  0.004 &  0.005 &  0.003 \\
			0 &  0.006 &  0.007 &  0.004 \\
			0 &  0.004 &  0.004 &  0.003 \\
			0 &  0.005 &  0.005 &  0.003 \\
        \end{bmatrix}.
    \end{equation}
\label{eq:modtopbot}
\end{subequations}

\begin{subequations}
	\begin{equation}
	\Vec{D_{top}}  = 
	\begin{bmatrix}
		 0.284 &  0.213 &  0.230 &  0.273 \\         
		 0.452 & -0.504 &  0.478 & -0.425 \\  
		-0.471 &  0.442 &  0.483 & -0.455 \\         
		 0.428 &  0.456 & -0.476 & -0.408 \\  
	\end{bmatrix}
	\pm
	\begin{bmatrix}
		0.003 &  0.003 &  0.004 &  0.003 \\
		0.006 &  0.006 &  0.007 &  0.006 \\
		0.006 &  0.006 &  0.007 &  0.006 \\
		0.006 &  0.006 &  0.006 &  0.006 \\
	\end{bmatrix},
	\end{equation}
	
	\begin{equation}
	\Vec{D_{bot}} = 
	\begin{bmatrix}
		 0.242 &  0.260 &  0.232 &  0.266 \\    
		-0.469 &  0.467 & -0.450 &  0.452 \\
		 0.448 & -0.451 & -0.471 &  0.473 \\   
		-0.429 & -0.446 &  0.437 &  0.437 \\
	\end{bmatrix}
	\pm
	\begin{bmatrix}
		0.003 &  0.003 &  0.003 &  0.003 \\
		0.006 &  0.006 &  0.006 &  0.006 \\
		0.006 &  0.006 &  0.006 &  0.006 \\
		0.006 &  0.006 &  0.006 &  0.006 \\
	\end{bmatrix}.
	\end{equation}
	\label{eq:demodtopbot}
\end{subequations}

From aforementioned demodulation matrices polarimetric accuracy was estimated to be better than $3\times 10^{-2}$. This also reflects in the difference between given and measured input Stokes parameters for calibration, as already depicted in \cref{fig:demodbal}.

\subsection{Observations of a sunspot} \label{ssc:obs}

The Sunspot NOAA-12706 was observed on April 26, 2018, using the instrument. Data acquisition, reduction procedure is same as described in \cref{ssc:dataacq}. Level-1 observation data were processed further to extract full Stokes parameters. Demodulation matrix given in \cref{eq:modtopbot} was applied to top and bottom beams separately. The two beams were aligned both spatially and, spectrally and combined to get Stokes-$I$, $Q/I$, $U/I$, $V/I$. It was observed that final Stokes-$Q/I$ profiles suffer from low frequency sinusoidal variation in spectral direction. These frequency components were removed by Fourier domain filtering.

The final step in the processing was to remove cross-talk due to instrumental polarization. $Q \rightarrow I$, $U \rightarrow I$ and $V \rightarrow I$ cross-talks were ignored as significant portion of the light would be unpolarized and the contribution from these terms would affect the total intensity in the order of 0.1\%. $I \rightarrow Q$, $I \rightarrow U$ and $I \rightarrow V$ manifest as offsets from zero in $Q/I$, $U/I$ and $V/I$ profiles. These are usually estimated by identifying the continuum level. But, considering the broad nature of spectral line, it was difficult to identify the continuum level. Hence, they were estimated by applying Fourier transform and calculating the contribution from zero frequency component. Results are given in \cref{eq:i2quv}.

\begin{subequations}
	\begin{equation}
		M_{I\rightarrow Q} = -0.095 \pm 0.004,
	\end{equation}
	\begin{equation}
		M_{I\rightarrow U} = -0.013 \pm 0.004,
	\end{equation}
	\begin{equation}
		M_{I\rightarrow V} = -0.007 \pm 0.003.
	\end{equation}
\label{eq:i2quv}
\end{subequations}

To estimate rest of the cross-talk terms, analysis given in Ref.~\cite{Sanchez_1992_ObsInterpAsymQUV} was followed with little modifications as outlined below. $Q\rightarrow V$, $U\rightarrow V$ were ignored as there were no parts of FoV with $Q >> V$ or $U >> V$. To estimate $V\rightarrow Q$ and $V\rightarrow U$ cross-talks, Stokes profiles of photospheric lines from umbral region were considered. They would likely to have strong $V$, and negligible $Q$ and $U$ components. From data, these $Q$ and $U$ signals above $3\sigma$ noise level were plotted against $V$ signal from the same spectral line position and straight lines were fitted to them. Slopes of the lines give $V \rightarrow Q$ and $V \rightarrow U$ cross-talk estimates. There was not enough signal to estimate $V\rightarrow Q$ cross-talk properly, however $V\rightarrow U$ cross-talk was estimated to be

\begin{equation}
	M_{V\rightarrow U} = -0.584 \pm 0.002.
\end{equation}

\begin{figure}
	\centering
	\includegraphics[width=150mm]{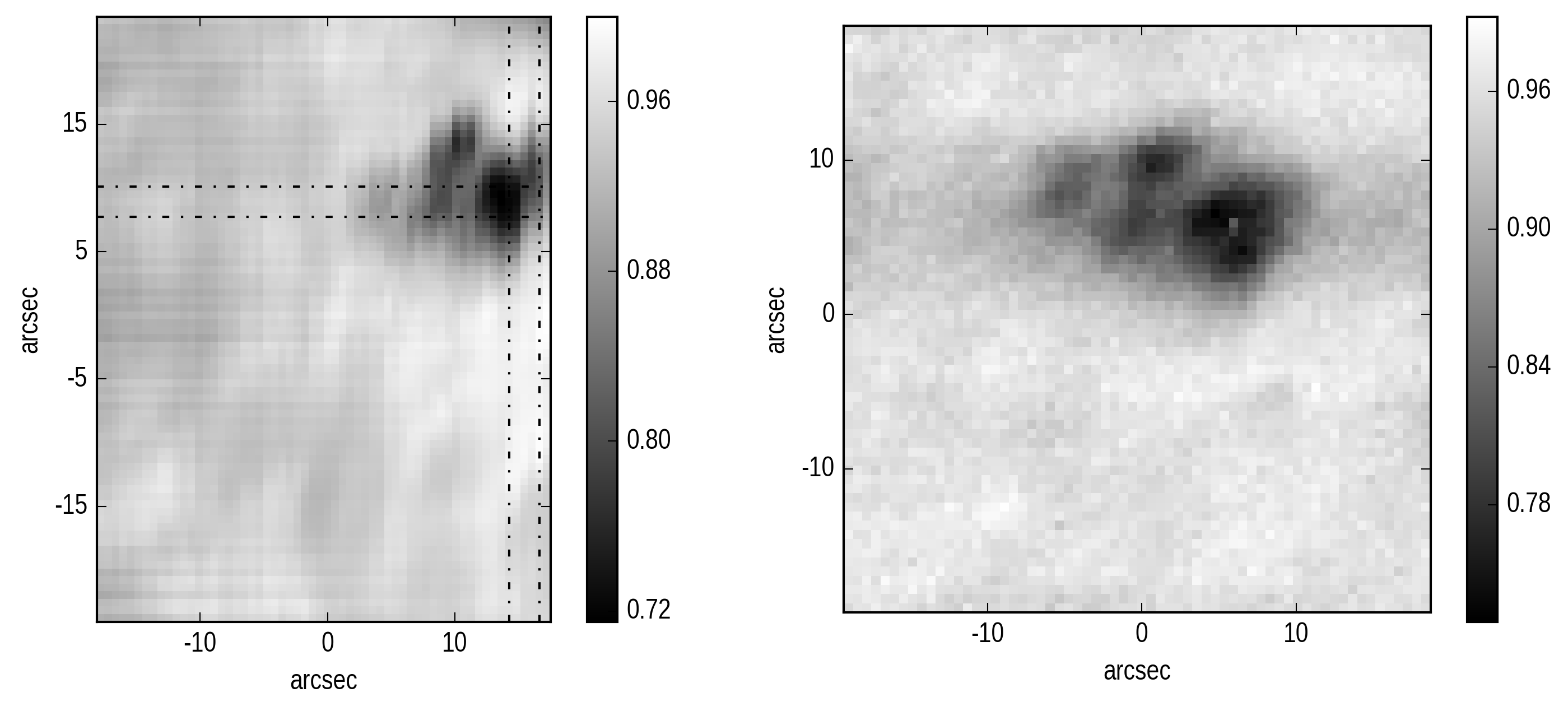}
	\caption{The left panel: Raster image in continuum with region of interest marked with dashed lines, the right panel: context image taken in continuum 600 nm/10 nm.}
	\label{fig:contim}
\end{figure}

\begin{figure}
	\centering
	\includegraphics[width=150mm]{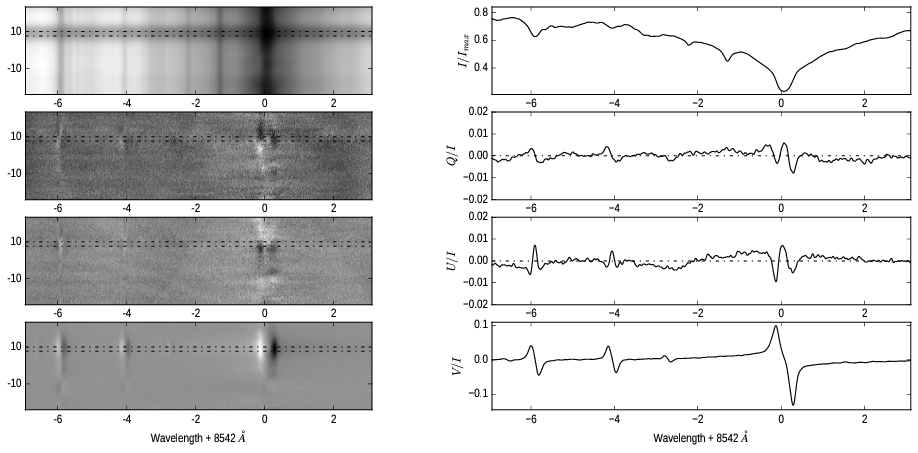}
	\caption{The left panel : Stokes spectra corresponding to vertical RoI marked in \cref{fig:contim}, right: Stokes profiles averaged over the spatial region marked between the horizontal dashed lines in the left panel.}
	\label{fig:spectraline}
\end{figure} 

After estimating the relevant cross-talks, corrections were applied and the resultants were saved as level-2 data. \cref{fig:contim} shows the raster image constructed with Stokes-I data, and context image. Integrated Stokes parameters of the RoI were plotted as shown in \cref{fig:spectraline}. These line profiles were smoothed with a Gaussian filter with $\sigma = \text{Slit width}/2 $. RMS value of $Q/I$ and $U/I$ from spectral continuum region were calculated. This is the polarimetric sensitivity i.e., the minimum amount of the polarization that can be detected. As a fraction of intensity it was calculated to be $3\times10^{-3}$/pixel which is reasonably good for the active region magnetic field measurements. These fluctuations could be stemming from residuals from flat fielding rather than photon noise.


\section{Summary} \label{sec:summary}
A new polarimeter has been designed and installed at the Tower-tunnel Telescope of KSO to probe the Chromosphere of the Sun, with 8542~\AA\ as the design wavelength. This polarimeter has inbuilt scanning mechanism to scan a RoI of the solar image, which emerged as a natural consequence of the instrument setup. The setup also has a provision for installing an auto-guiding system. Data reduction procedures have been developed in Python programming language which correct for bias counts, gain table variations of the pixels (flat fielding), polarimeter response (polarimetric calibration) and instrumental polarization. Sample Stokes images and profiles of a sunspot group observed using this polarimeter are presented. Further analysis of the data is under way.

\begin{appendices}
\crefalias{section}{appendix}
\section{Instrumental polarization model and simulation} \label{app:instrupol}
Instrumental polarization introduced by a mirror depends on the angles of incidence and the complex refractive index of the coating. In order to estimate the instrumental polarization of a series of mirrors, transformations of the local coordinate system also should be known. Refractive index of the unprotected Aluminium coating for 8542~\AA\ is taken from existing literature \cite{McPeak_2015_AlRI}.

The direction cosines (DC) of the vectors relevant to get the incident angles are calculated. The Coelostat model uses following convention for the coordinate system and angles.

\begin{itemize}
	\item Center of the Coelostat M2 mirror is the origin.
	\item West is positive $X$
	\item Zenith is positive $Y$
	\item North is positive $Z$
	\item Declination ($\delta$) is positive along the North
	\item Latitude ($\phi$) is positive for the South
	\item Hour angle (HA) is positive along the West
\end{itemize}

Parameters in the model are

\begin{itemize}
	\item Declination varies from $-23.5 \degree$ to $23.5 \degree$
	\item HA varies from $-90 \degree$  to $90 \degree$ 
	\item Latitude is fixed at $-10.24 \degree$
	\item $X$ distance between M1 and M2 ($a$) is a free parameter and it is fixed at $\pm830 mm$
	\item $Y$ distance between M1 and M2 ($c$) is a free parameter and it is fixed at $740 mm$
	\item $Z$ distance between M1 and M2 ($b$) depends on $a$, $c$ and $DEC$ 
\end{itemize}

DC of the Sun's position and subsequently the incidence angle on the M1 are given by

\begin{subequations}
	\begin{equation}
	\begin{aligned}
	\Vec{P_{Sun}} =
	&(\sin{HA} \cos{\delta} \\
	&\cos{HA}\cos{\delta}\cos{\phi} - \sin{\delta}\sin{\phi} \\
	&\cos{HA}\cos{\delta}\sin{\phi} + \sin{\delta}\cos{\phi}),
	\end{aligned}
	\end{equation}
	\begin{equation}
	\begin{aligned}
	\Vec{i_{M1}} &= - \Vec{P_{Sun}}.
	\end{aligned}
	\end{equation}
\end{subequations}

DC of the rotation axis of the primary mirror i.e., polar axis, are given by

\begin{equation}
\Vec{P_{Pol}} = (0, -\sin{\phi}, \cos{\phi}).
\end{equation}

The reflected beam must be towards the secondary mirror and its DC are given by

\begin{equation}
\Vec{r_{M1}} = \frac{(a, c, b)}{\sqrt{a^2+b^2+c^2}}.
\end{equation}
DC of the M1 normal will be

\begin{equation}
\Vec{n_{M1}} = \frac{\Vec{r_{M1}}-\Vec{i_{M1}}}{|\Vec{r_{M1}}-\Vec{i_{M1}}|}.
\end{equation}

This must be normal to the polar axis and solving \cref{eq:nprimppol} results in the value for $b$ given in \cref{eq:bdist}.

\begin{subequations}
	\begin{equation}
	\Vec{n_{M1}}.\Vec{P_{Pol}} = 0,
	\label{eq:nprimppol}
	\end{equation}
	\begin{equation}
	b = B + \frac{\delta}{|\delta|}\frac{\sqrt{B^2-4AC}}{2A},
	\label{eq:bdist}
	\end{equation}
\end{subequations}

where $A, B, C$ are given by

\begin{align*}
A &= \cos^2{\phi} - (\Vec{i_{M1}}.\Vec{P_{Pol}})^2, \\
B &= -2c\sin{\phi}\cos{\phi}, \\
C &= c^2\sin^2{\phi} - (a^2+c^2)(\Vec{i_{M1}}.\Vec{P_{Pol}})^2. \\
\end{align*}

Now, consider the polarimeter side. Position and orientation of the M3 and MSM are fixed. So, the DC for the beam entering the polarimeter i.e., the beam which is reflected from the MSM, and the DC for the beam incident on the MSM are given by

\begin{subequations}
	\begin{equation}
	\Vec{r_{MSM}} = (-1, 0, 0),
	\end{equation}
	\begin{equation}
	\Vec{i_{MSM}} = (\sin{0.4\degree}, 0, \cos{0.4\degree}),
	\end{equation}
	\begin{equation}
	\Vec{n_{MSM}} = \frac{\Vec{r_{MSM}}-\Vec{i_{MSM}}}{|\Vec{r_{MSM}}-\Vec{i_{MSM}}|}.
	\end{equation}        
\end{subequations}

DC for M3 normal, incident and reflections are given by

\begin{subequations}
	\begin{equation}
	\Vec{n_{M3}} = (0,\cos{45\degree}, \sin{45\degree}),
	\end{equation}
	\begin{equation}
	\Vec{r_{M3}} = \Vec{i_{MSM}},
	\end{equation}
	\begin{equation}
	\Vec{i_{M3}} = \Vec{r_{M3}} - 2(\Vec{r_{M3}}.\Vec{n_{M3}})\Vec{n_{M3}}.
	\end{equation}        
\end{subequations}

From the above equations, DC of M2 normal as well as incidence and reflection are given by

\begin{subequations}
	\begin{equation}
	\Vec{i_{M2}} = \Vec{r_{M1}},
	\end{equation}
	\begin{equation}
	\Vec{r_{M2}} = \Vec{i_{M3}},
	\end{equation}
	\begin{equation}
	\Vec{n_{M2}} = \frac{\Vec{r_{M2}}-\Vec{i_{M2}}}{|\Vec{r_{M2}}-\Vec{i_{M2}}|}.
	\end{equation}        
\end{subequations}

The electric field vector can be decomposed to have two orthogonal components: parallel ($p$) and perpendicular ($s$) to the plane of incidence and reflection. Together, the direction of propagation, $s$ direction and $p$ direction form a right handed coordinate system. DC for the $s$ polarized components corresponding to each mirror are

\begin{subequations}
	\begin{equation}
	\Vec{s_{M1}} = 
	\frac{\Vec{i_{M1}}\times\Vec{n_{M1}}}{|\Vec{i_{M1}}\times\Vec{n_{M1}}|},
	\end{equation}
	\begin{equation}
	\Vec{s_{M2}} = 
	\frac{\Vec{i_{M2}}\times\Vec{n_{M2}}}{|\Vec{i_{M2}}\times\Vec{n_{M2}}|},
	\end{equation}
	\begin{equation}
	\Vec{s_{M3}} = 
	\frac{\Vec{i_{M3}}\times\Vec{n_{M3}}}{|\Vec{i_{M3}}\times\Vec{n_{M3}}|}.
	\end{equation}
	\begin{equation}
	\Vec{s_{MSM}} = (0, -1, 0)
	\end{equation}
\end{subequations}

The three angles for three coordinate rotations, considering the signs, are be given by

\begin{subequations}
	\begin{equation}
	\theta_{M1-M2} = 
	\frac{(\Vec{s_{M1}}\times\Vec{s_{M2}}).\Vec{r_{M1}}}
	{|(\Vec{s_{M1}}\times\Vec{s_{M2}}).\Vec{r_{M1}}|}
	\cos^{-1}(\Vec{s_{M1}}.\Vec{s_{M2}}),
	\end{equation}
	\begin{equation}
	\theta_{M2-M3} = 
	\frac{(\Vec{s_{M2}}\times\Vec{s_{M3}}).\Vec{r_{M2}}}
	{|(\Vec{s_{M2}}\times\Vec{s_{M3}}).\Vec{r_{M2}}|}
	\cos^{-1}(\Vec{s_{M2}}.\Vec{s_{M3}}),
	\end{equation}
	\begin{equation}
	\theta_{M3-MSM} = 89.6\degree.
	\end{equation}
\end{subequations}

As the complete information about the mirrors' positions and orientations are known, the system Mueller matrix can be constructed as given in \cref{eq:analinstrupol}. This is a function of $HA$ and $\delta$. Here, $\Vec{M}$ represents Mueller matrix of a mirror and $\Vec{R}$ represents Mueller matrix for rotation. 

\begin{subequations}
	\begin{equation}
	\Vec{M_{time-var}} =
	\Vec{R}(\theta_{M2-M3})\Vec{M_{M2}}
	\Vec{R}(\theta_{M1-M2})\Vec{M_{M1}},
	\end{equation}
	\begin{equation}
	\Vec{M_{time-inv}} = 
	\Vec{M_{MSM}}
	\Vec{R}(\theta_{M3-MSM})\Vec{M_{M3}},
	\end{equation}
	\begin{equation}
	\Vec{M_{instru}} = \Vec{M_{time-inv}}\Vec{M_{time-var}}.
	\end{equation}
	\label{eq:analinstrupol}
\end{subequations}

The system is simulated in the \textsc{Zemax} non-sequential mode. PyZDDE \cite{PyZDDE} - a Python-\textsc{Zemax} interface is used to interactively change the parameters and perform ray tracing for this dynamic system. Instrumental polarization of the system is calculated by giving six inputs ($\pm Q, \pm U, \pm V$, one at a time) and performing polarimetry at the output. The simulation is done for the time varying and time independent parts separately for the sake of comparison.

\section{Creating the master flat} \label{app:flatmaster}
Apart from the process described in Ref.~\cite{Wohl_2002_PrecRedSolSpecCCD}, an additional process was followed to reduce the effect of the fringes. Fringe flat frames obtained as a part of acquisition were used in the process to create the master flat. They were acquired by capturing continuum with no spectral line, close to 8542 \AA. They were averaged and master dark was subtracted from it. The intensity gradient across the fringe flat was computed and corrected.

\begin{figure}
	\centering
	\includegraphics[width=150mm]{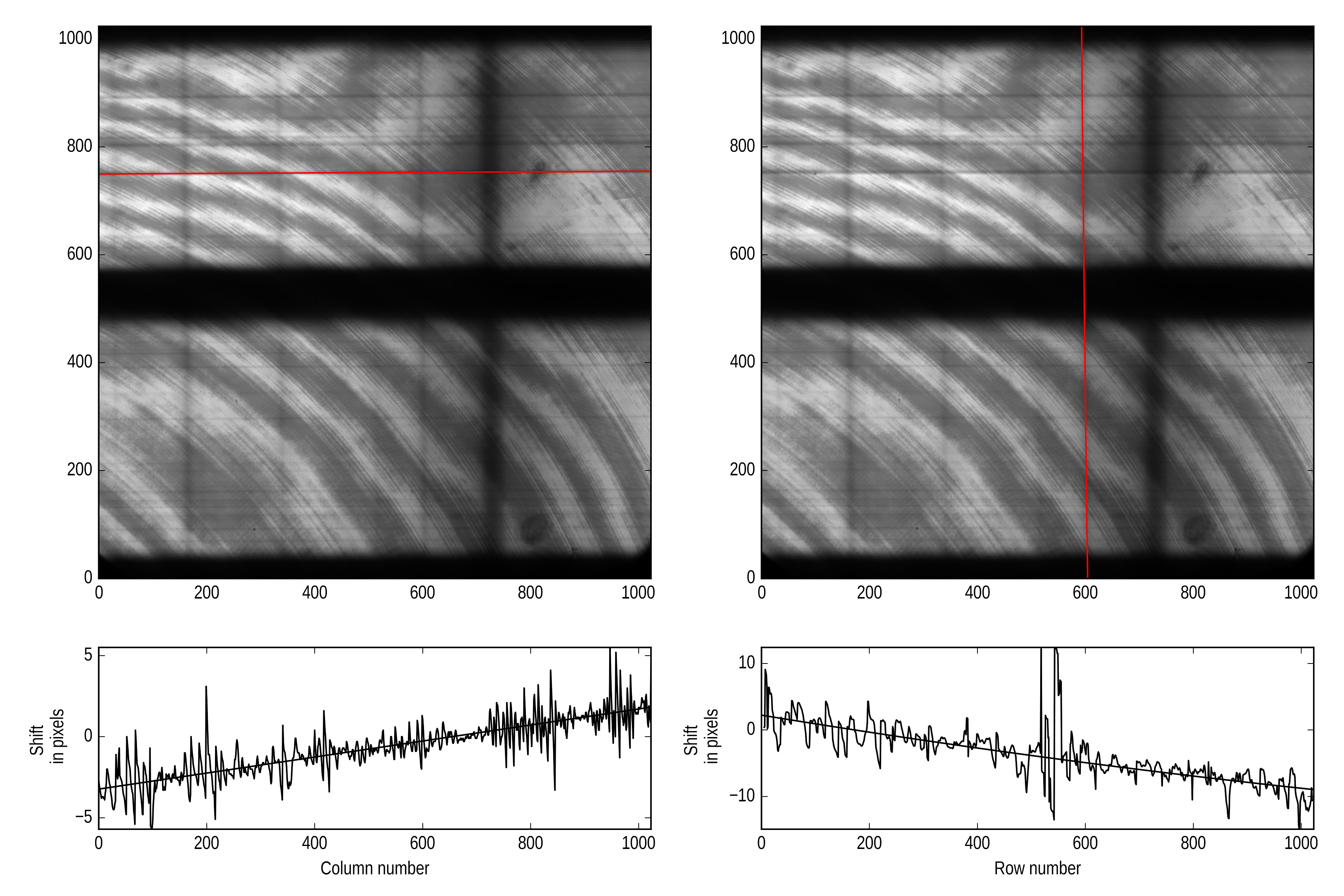}
	\caption{Top-left: slit feature trace, bottom-left: column shift values for each row, top-right: line profile trace, bottom-right: row shift values for each column.}
	\label{fig:rowcols}
\end{figure}

\begin{figure}
	\centering
	\includegraphics[width=150mm]{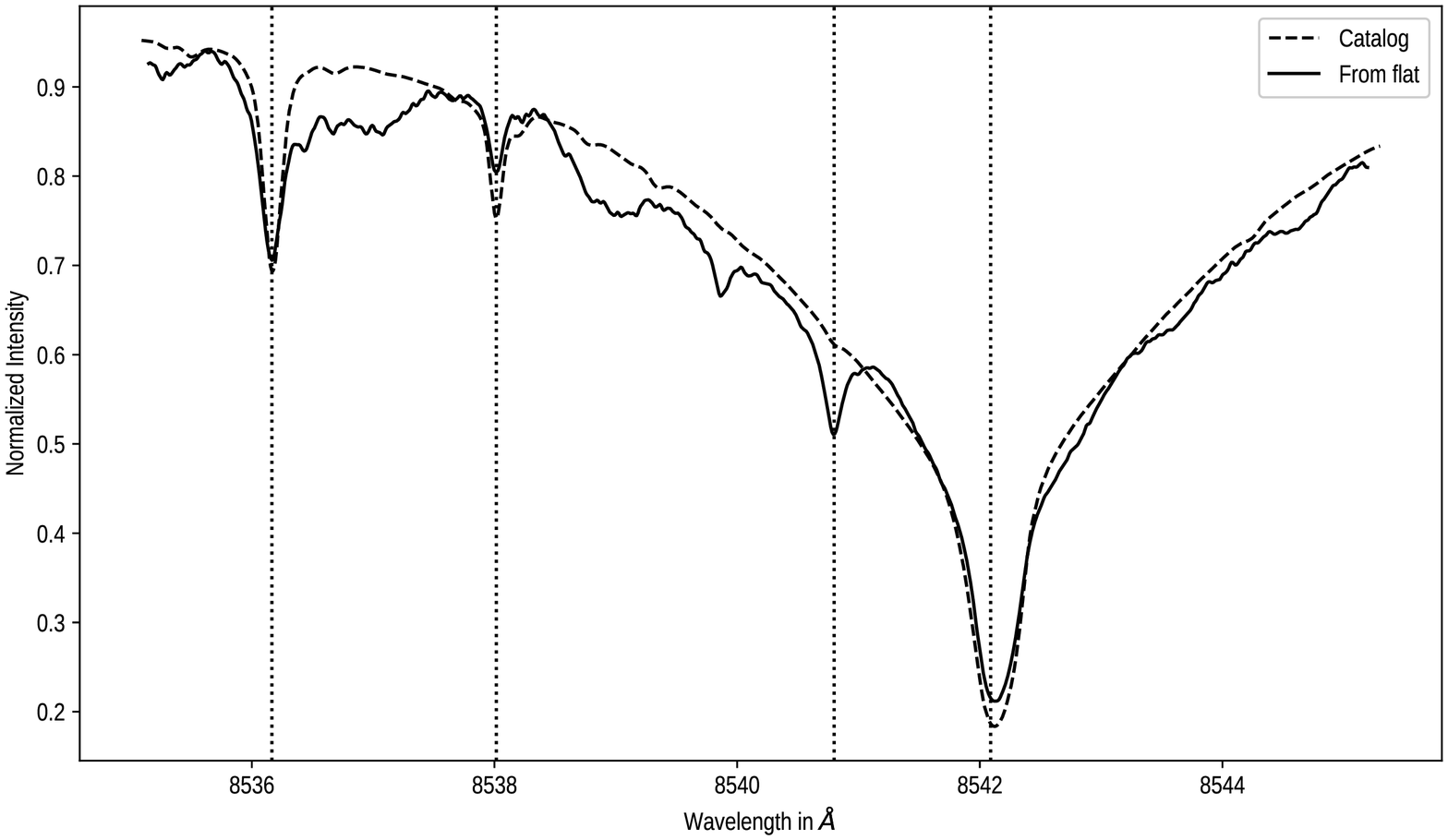}
	\caption{Spectrum obtained from the disk center over-plotted  with the one from catalog.}
	\label{fig:catlins}
\end{figure}

Mean main flat must be aligned in such a way that the spectral and spatial axes are along reference axes. This called for three types of correction: 1) X inclination, 2) Y inclination, 3) Y curvature. They are caused due a combination of distortion caused by optics, slit and detector reference axis not being parallel, and slit and grating grooves not being parallel. As they are very small in amount compared to the detector size, they can be corrected by shifting columns and rows respectively. To correct X inclination, a slit pattern with good contrast was selected and traced across the detector. A line was fitted to this trace and each column was shifted by the amount given by that linear equation. Before correcting for the Y inclination, mean flat was divided by fringe flat to which the same column shifts were applied. This resulted in reduced fringe contrast. Then, atmospheric line at 8540.7 \AA\ was traced and a quadratic equation was fitted to it to get both Y inclination and curvature. All the rows were shifted by the amounts given by this equation. Row and column shift values were saved so that they can be used in further reduction. Correcting the mean main flat for the inclinations resulted in the spectral line being aligned vertically. \cref{fig:rowcols} illustrates tracing of slit and spectral line features, and corresponding column and row shifts.

The median spectral line profile was calculated using the central parts of both top and bottom beams. This profile is smoothed by a Gaussian filter to reduce spurious pixel to pixel variation. Each row of the aligned mean flat was divided by smoothed line profile, thus creating the master flat. \cref{fig:flats} illustrates row and column shifts, raw flat file and the master flat. Spectral line profile is also saved to serve as a template, as it was obtained from averaged disk center. It would be used to perform wavelength calibration.
\end{appendices}

\renewcommand{\abstractname}{Acknowledgements}
\begin{abstract}
We thank Indian Institute of Astrophysics mechanical and electronics division personnel for their contribution in design and development of the instrument. We also thank KSO mechanical and electrical division, and observing personnel for their support in installation and testing. We thank the chair and members of ``The Sun and solar system" division for their support in facilitating the observing time. We thank the two anonymous referees for their critical review and insightful comments that assisted in improving this paper. 
\end{abstract}

\bibliographystyle{ieeetr}   
\bibliography{references}   
\end{document}